\documentstyle[prl,epsfig,twocolumn,aps]{revtex}

\begin{document}

\draft

\title{$1D$ Aging}
\author{L.R.~Fontes}
\address{Instituto de Matem\'atica e Estat\'\i stica, Universidade 
         de S\~ao Paulo, 05315-970 S\~ao Paulo, Brasil}
\author{M.~Isopi}
\address{Dipartimento di Matematica, Universit\`a di Bari,
70125 Bari, Italia}
\author{C.M.~Newman}
\address{Courant Institute of Mathematical Sciences,
New York University, New York, NY 10012}
\author{D.L.~Stein}
\address{Departments of Physics and Mathematics, University of Arizona,
Tucson, AZ 85721}

\maketitle

\begin{abstract}
We derive exact expressions for a number of aging functions that are
scaling limits of non-equilibrium 
correlations, $R(t_w, t_w+t)$ as $t_w \to
\infty$, $t/t_w \to \theta$, in the $1D$ homogenous $q$-state Potts model
for all $q$ with $T=0$ dynamics following a quench from $T = \infty$. One
such quantity is $\langle \vec{\sigma}_0(t_w) \cdot \vec{\sigma}_n(t_w+t)
\rangle$ when $n/\sqrt{t_w} \to z$. Exact, closed-form expressions
are also obtained when one or more interludes of $T = \infty$ dynamics
occur.  Our derivations express the scaling limit via coalescing Brownian
paths and a ``Brownian space-time spanning tree,'' which also yields other
aging functions, such as the persistence probability of no spin flip at $0$
between $t_w$ and $t_w+t$.

\end{abstract}


Aging and related memory effects are a fundamental feature of
nonequilibrium dynamics \cite{review}.  First observed in mechanical
properties of glassy polymers and other amorphous materials \cite{Struik},
it was found to be a central property of spin glass dynamics as well
\cite{LSNB83}.  Although originally thought to be a distinguishing feature
of disordered systems, aging is now known to occur in both homogeneous and
disordered systems following a quench to low temperature $T$ (though debate
persists over whether aging differs qualitatively in different systems
\cite{VDAHB}).  Although many models and
mechanisms for aging have been proposed, few exact (mostly, but not
exclusively, for the $1D$ Ising chain) or rigorous results exist
\cite{FKR97,Bray97,GL00,benarous,FIN00}.

In this paper we present a general approach to aging in $1D$ discrete spin
models (and equivalent systems, such as reaction-diffusion or voter
models), in the continuum space-time scaling limit (e.g., lattice spacing
$a\to 0$, with time scaled by $a^2$).  We focus here on exact solutions for
the scaling limit of the entire dynamical process, and thence for aging
functions, in homogeneous ferromagnetic $q$-state Potts models (where $q=2$
is the Ising model), but the approach is also applicable to inhomogeneous
systems, as in \cite{FIN00}.  The method uses earlier work
by Arratia \cite{Arratia83} to express the scaling limit via a ``Brownian
space-time spanning tree'', depicting the histories of coalescing Brownian
particles in one dimension starting from all possible locations and times.

In a typical aging experiment, a system is rapidly quenched from high to low
$T$.  After a time $t_w$ following the quench, an external
parameter (e.g., temperature or external field) is changed.  The response
$R(t_w, t_w+t)$ of the system (e.g., decay of thermoremanent magnetization)
is then measured at time $t_w+t$.  Aging can also be observed without a
sudden parameter change; e.g., in the out-of-phase component
$\chi''$ of the ac susceptibility \cite{review}.  In either case, the
essence of aging is that as $t_w\to\infty$,
$t\to\infty$, the response depends only on the ratio $t/t_w$:
\begin{equation}
\label{eq:response}
\lim_{\scriptstyle{t\to\infty},
\scriptstyle{t_w\to\infty}\atop\scriptstyle{t/t_w\to\theta}}R(t_w, t_w+t)=
{\cal R} (\theta).
\end{equation}
(Other scaling forms
are discussed in \cite{RMB}.)  Equilibrium responses are 
time-translation-invariant, so aging is a non-equilibrium,
history-dependent phenomenon.

There already exist a few exact results for related quantities measuring
coarsening \cite{Bray94} or persistence \cite{NS99}.  Exact results for
persistence exponents (and fraction of persistent spins) in $1D$ Potts
models appear in \cite{DHP96}, and results on coarsening quantities (such
as domain size distributions) appear in \cite{SM95,KB-N97,B-AM00}. 
There are fewer exact results for aging quantities.
An exception is the 
well understood Ising chain, for which two-time
correlations have been derived
\cite{Bray97} (see also \cite{GL00}).  
However, the methods used seem specialized to the Ising case.  For general
Potts models, exact results have been obtained only for $F_q(t_w, t_w+t)$,
the probability of no spin flip at the origin (in $1D$) between $t_w$ and
$t_w+t$.  This was analyzed in \cite{DHP96} for a semi-infinite chain and,
for $q=\infty$, in \cite{FKR97} on the full $1D$ lattice.  In the following
sections we present our general method and compute exact results for a
variety of aging quantities in the continuum scaling limit, including as
special cases rederivations of the results of \cite{FKR97,Bray97}.

{\it Preliminaries.\/} Consider the homogeneous $q$-state ferromagnetic
Potts model on the $1D$ integer lattice, where the Potts spin variables
$\sigma_n$, for $-\infty < n < \infty$, can take the values $1,\ldots,q$.
We study $T=0$ dynamics following a quench from $T=\infty$ --- i.e., in the
initial $\sigma(0)$, each site independently takes a random value uniformly
from $1,\ldots,q$.  (For $q=\infty$, each site takes its own unique value.)
The standardly used (as in \cite{DHP96}) continuous time $T=0$ dynamics,
that of the $1D$ voter model (see, e.g., \cite{Liggett}), is given by
independent Poisson ``clock'' processes at each site $n$, all of rate one,
indicating when a flip at $n$ is considered.  When the clock at $n$ rings,
$\sigma_n$ takes the value of one neighbor, chosen by a fair coin toss
(regardless of whether it previously agreed with either or both neighbors).
There are other natural $T=0$ dynamics, but we defer their analysis to a
later paper.

In studying this evolution, it is convenient to use a
well-known mapping to a $1D$ reaction-diffusion system of ``kinks''
\cite{Schwartz,R85,AF90}.  A kink corresponds to a site $n+1/2$ in the {\it
dual\/} lattice where $\sigma_n\ne\sigma_{n+1}$.  The initial configuration
is a random arrangement of kinks (with density
$(q-1)/q$) that subsequently execute $1D$ random walks.  For the Ising
model ($q=2$), the walks are purely annihilating, while for $q=\infty$ they
are purely coalescing; for other $q$ both annihilation and coalescence
occur.

At time $t_w$, there will be a characteristic distribution, depending on
$q$, of walkers (i.e., kinks). One aging quantity is the persistence
probability $F_q(t_w,t_w+t)$, mentioned earlier, of no flip at the origin
between $t_w$ and $t_w+t$.  A related quantity is $G_q(0,0;t_w,t_w+t) =
\langle \delta_{\sigma_0(t),\sigma_0(t_w+t)}\rangle$, the probability $P$
that $\sigma_0(t_w+t) = \sigma_0(t)$ (regardless of intervening flips).
More generally, $G_q(m,n;s,s')$ is $P(\sigma_n(s') = \sigma_m(s))$, and the
spin-spin correlation $C_q = [q/(q-1)](G_q - 1/q)$, which is $\langle
\vec{\sigma}_m(s') \cdot \vec{\sigma}_n(s) \rangle$ in the ``tetrahedral''
representation of Potts spins.  Other aging quantities will be discussed
later.

\begin{figure}
\vspace{-0.6in}
\centerline{\epsfig{file=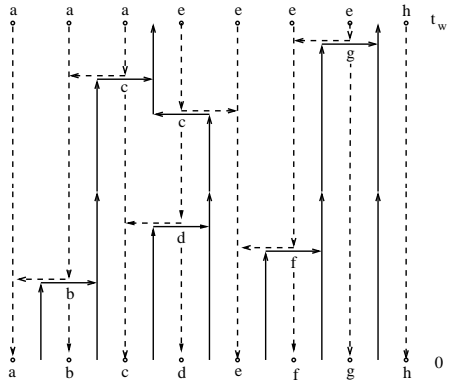,width=2.5in}}
\vspace{-0.4in}
\caption{An eight-site lattice for the $q=\infty$ model.
At $t=0$ all spins have distinct values, labeled a-h.  Both forward in $t$
coalescing walks representing domain boundary motion (solid lines), and
backward in $t$ coalescing walks representing ``ancestry'' (dashed lines)
are shown; horizontal segments are at times of Poisson clock rings.  This
diagram can be used for any $q$; e.g., for $q=2$, abdgh might all be $+1$
and cef $-1$.}
\label{fig:coalescence}
\end{figure}

{\it Coalescing Brownian paths and spanning trees.\/} The coalescing random
walks of kinks correspond exactly to the motion of the boundaries between
clusters of like spins for the $q=\infty$ model. Whether
$q=\infty$ or not, to determine $\sigma_m(t_w)$, 
it is natural to trace backward in time to see successively from which
neighbor the spin value came as various clocks rang. This leads to a
dual process \cite{Liggett} of coalescing random walks on the {\it
original\/} lattice such that all sites $n$ whose (backward in time)
walkers have coalesced and are located at $\ell$ at time zero have
$\sigma_n(t_w) = \sigma_\ell(0)$ (see Fig.~1).

Thus, for $q=\infty$, the equal-time $G_\infty (m,n;t_w,t_w)$
$= P(\sigma_m(t_w) = \sigma_n(t_w))$ is just the probability that two
{\it independent\/} (backward in time) random walks starting at $m$ and $n$ 
meet within time $t_w$.  But the dual process also works for unequal
times, so $G_\infty (m,n;t_w,t_w+t)$ equals the probability that two
walkers, one starting at $m$ and the other starting $t$
units of time ``earlier'' at $n$ meet and hence
coalesce between times $0$ and
$t_w$.

Not surprisingly, in the scaling limit, the walkers (both forward and
backward) become particles doing Brownian motion. Concretely, when $t_w \to
\infty$, one rescales the lattice by $a=1/\sqrt{t_w}$ and time by $a^2$, so
that backward walkers starting from sites $0$ and $\sqrt{t_w}z$ become
Brownian particles, one from $0$ and the other starting $\theta = \lim
t/t_w$ units of time ``earlier'' from $z$. This will be used below.

More surprisingly, a scaling limit is valid not just for a few
walkers, but simultaneously for walkers starting 
from {\it every\/} lattice site at $t=0$ \cite{Arratia83}.  
The limit essentially has Brownian particles
starting from every point on the continuous line at $t=0$, but for any
$t>0$, coalescing has reduced them to a discrete set.

An extended limit, useful for understanding aging of persistence
quantities, includes all starting times and simultaneously the (backward in
time) dual particles with all {\it their\/} starting times. The collection
of all forward (resp., backward) space-time paths forms a spanning tree of
continuum space-time in the sense of \cite{ABNW}.

{\it Spin-spin correlation.\/} As in the previous section, we express
$G_\infty(0,\sqrt{t_w}z;t_w,t_w+t)$ in the scaling limit via coalescing
dual Brownian paths. The limit of $G_\infty$, denoted $g(z,\theta) =
g_\infty(z,\theta)$, is the probability that the backward (i.e., dual)
Brownian paths starting at the space-time points $(z,1+\theta)$ and $(0,1)$
coalesce before time $0$. Denote the location at time $t-s$ of the backward
Brownian path starting at a generic $(x,t)$ by $\tilde B_{x,t}(s), s\geq0$.
Conditioning on the value $x$ of $\tilde B_{z,1+\theta}(\theta)$, we have
 
\begin{eqnarray}
\label{eq:corr1}
g(z,\theta)=\frac{1}{\sqrt{2\pi\theta}} \int_{-\infty}^\infty dx\, 
e^{-(x-z)^2/2\theta} g(x)\,,
\end{eqnarray}
with $g(x) = P(A_x)$, where $A_x$ is the event that $\tilde B_{0,1}$ and
$\tilde B_{x,1}$ coalesce at some $s \in [0,1]$.

Note from~(\ref{eq:corr1}) that $g(z,\theta)$ satisfies the
heat equation, $\partial g / \partial \theta = (1/2) \partial^2 g /
\partial z^2$ (see, e.g., \cite{Bray97} for a
corresponding result in the Ising case), 
with $g(z,0) = P(A_z)$.  Since $A_x$ is the event that
$\tilde B_{0,1}(s)-\tilde B_{x,1}(s)=0$ for some $s\in[0,1]$, and the
difference of two independent (before coalescing) Brownian motions of rate
$1$ is a Brownian motion of rate $2$, we can rewrite $P(A_x)$ as
$P(B_x(s)=0$ for some $s\in[0,1])$ $=1-P(B_x \neq 0$ during $[0,1])$, where
$B_x(s)$ is a Brownian motion starting at $x$ of rate $2$.  By a standard
argument using the Reflection Principle \cite{Feller} and the symmetry in
$x$, the latter probability equals $P(B_{|x|}(1)>0)-P(B_{-|x|}(1)>0)$, and
this equals $P(B_{0}(1)>-|x|)-P(B_{0}(1)>|x|)=2P(0<B_0(1)<|x|)
=\phi(|x|/\sqrt 2),$ where $\phi(x)=\sqrt{2/\pi}\int_0^x dt\, e^{-t^2/2}$.

Substituting in~(\ref{eq:corr1}) and rewriting again, we find that
$1-g(z,\theta)$ equals $(2\pi\theta)^{-1/2}[h(z)+h(-z)]$, where
$h(z)=\int_{-z}^\infty
dx\,e^{-x^2/2\theta}\phi\left((x+z)/\sqrt2\,\right)$.
After further analysis,
\begin{eqnarray}
\label{eq:corr4}
g(z,\theta)=\psi(|z|/\sqrt{2+\theta},\sqrt{2/\theta}), 
\end{eqnarray}
where $\psi(a,b)=\sqrt{\frac2\pi}\int_a^\infty dt e^{-t^2/2}\phi(bt)$,
and finally
\begin{eqnarray}
\label{eq:corr5}
g(z,\theta)=\frac2\pi\int_0^{\sqrt{2/\theta}}dt\,
\frac{e^{-z^2(1+t^2)/(2(2+\theta))}}{1+t^2}.
\end{eqnarray}

Eqns.~(\ref{eq:corr4})-(\ref{eq:corr5}) simplify in particular cases, e.g.~
\begin{eqnarray}
\label{eq:corr6}
&g(0,\theta)=\frac2\pi\mbox{arctg}\sqrt{2/\theta},&\\
\label{eq:corr7} 
&g(z,0)=1-\phi(|z|/\sqrt2),\,\,
g(z,2)=\frac12[1-\phi^2(|z|/2)]\, .& 
\end{eqnarray}
$g(0,\theta)$ gives the scaling limit probability 
that $\sigma_m(t_w+t)=\sigma_m(t_w)$, regardless of flips 
during $(t_w, t_w+t)$.  $g(z,0)$ is the scaling limit equal-time
two-point correlation function and its exact formula in (\ref{eq:corr7}) is
implicit or explicit in earlier work on inter-particle distributions
\cite{Arratia83,DZ96,DBA88}.  

The $q < \infty$ case of $g_q(z,\theta)$ is simply related to the $q =
\infty$ case just discussed.  Clearly, $\sigma_{\sqrt{t_w}z} (t_w+t) =
\sigma_0(t_w)$ (in the scaling limit) if the backward Brownian paths
starting at the space-time points $(0,1)$ and $(z,1+\theta)$ coalesce
before time $0$. If not, there is still a $1/q$ probability that those
paths end at time $0$ on sites with the same spin value.  Hence
$g_q(z,\theta)= g(z,\theta) + \frac1q (1-g(z,\theta))$ and so
$c_q(z,\theta)$ is
\begin{eqnarray}
\label{eq:corr8}
\lim_{\scriptstyle{t\to\infty},
\scriptstyle{t_w\to\infty}\atop\scriptstyle{t/t_w\to\theta,\,
m/\sqrt{t_w} \to z}} C_q(0,m;t_w,t_w+t)=g(z,\theta),
\end{eqnarray}
which in particular {\it does not depend on $q$\/}.  Thus our exact results
(\ref{eq:corr1}) and (\ref{eq:corr6}) reproduce, as a special case, the
known Ising result for $c_2(z,\theta)$ (see \cite{Bray97,GL00}).

{\it $T=\infty$ interludes.\/} We now modify the dynamics by
inserting 
an interval of duration $\Delta$ with $T=\infty$ dynamics. 
I.e., when the clock at $n$ rings during
such an interlude (which it still does at rate one), $\sigma_n$ chooses a
value uniformly at random from $\{1,\dots,q\}$ (including the 
previous value);
for $q=\infty$, the new value is chosen to be
distinct from all other sites at that time.  The entire interlude is
inserted at a time $t_I$ (of the unmodified dynamics), which may be either
in $(0,t_w)$ or $(t_w,t_w + t)$.

Using the backward random walks of the (unmodified)
$q=\infty$ model,
to analyze the aging quantities $G^I_\infty = C^I_\infty$
for the modified dynamics, we consider walkers starting
from $(m,t_w)$ and $(n, t_w + t)$ and the events $A$ that
they coalesce at a time $\tau > 0$ and $B$, that 
for $\tau < t_I$, they pass
through the $T=\infty$ interlude {\it with no clock ring\/}.
Then $G^I_\infty = P(A \cap B)$. Furthermore, by the nature of 
the $T=\infty$ dynamics for finite $q$, it is clear that
$G^I_q = G^I_\infty + \frac1q (1 - G^I_\infty)$, which yields
$C^I_q = C^I_\infty = G^I_\infty$ for all $q$. 

Next note that the probability that a walker
passes through the interlude with no clock ring is $e^{-\Delta}$.  For
$t_I$ in $(t_w,t_w + t)$, $P(A)$ is just the unmodified $G_\infty$, so 
$C^I_q = e^{-\Delta} C_q = e^{-\Delta} G_\infty$
($=e^{-\Delta} g$ in the scaling limit); this is independent of
the location of $t_I$ within $(t_w,t_w + t)$.

For $t_I < t_w$, we partition $A$ into $A_1$, where $ t_I > \tau $ and
there are two independent walkers during the interlude (so $P(B)=
e^{-2\Delta}$) and the remainder $A_2$.  $A_2$ is the event that
coalescence occurs at $\tau > t_I$, so $P(A_2) = G_\infty (m, n; t_w - t_I,
t_w + t - t_I)$ and $C^I_q = P(A_2) + e^{-2\Delta} (P(A)-P(A_2))$.  Taking
the limit 
with $\Delta$ fixed, $(t_w -t_I)/t_w \to \rho > 0$,
$t/t_w \to \theta > 0$ and $(n-m)/\sqrt{t_w} \to z$, 
the spin-spin correlation with one $\Delta$-interlude is (see Fig.~2):
\begin{equation}
\label{eq:interlude}
c_q^I(z, \theta, \rho) = g(z,\theta/\rho) + e^{-2\Delta} [g(z,\theta) -
g(z,\theta/\rho)].
\end{equation}

\begin{figure}
\vspace{-1.35in}
\centerline{\epsfig{file=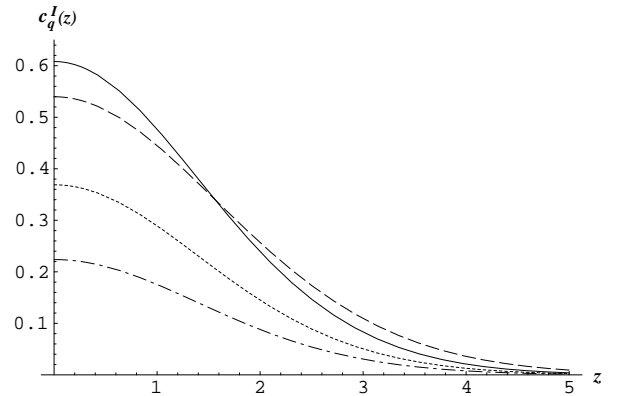,width=3.5in}}
\vspace{-1.2in}
\caption{One-interlude $c_q^I(z)$ curves with $\theta=1$, $\Delta = \frac12$:
solid ($=c_q(z,1)=g(z,1)$) for $t_I=0$, dashed for 
$t_I=\frac12 t_w$, dot-dashed for $t_I=t_w-$ , and dotted
for $t_I$ in $(t_w, t_w+t)$.}
\label{fig:correlation}
\end{figure}

Similar arguments for multiple $\Delta_j$-interludes at times $(1 - \rho_j)
t_w$ ($0<x\rho_1<\dots<\rho_\ell<1$) and ${\tilde \Delta}_j$-interludes at
times $(1+{\tilde \rho}_k)t_w$ ($0< {\tilde \rho}_k < \theta$) yield for
$c_q^I$:
\begin{equation}
\label{eq:interludes}
e^{-{\tilde \Delta}} \sum_{j=0}^\ell \exp(-2 \sum_{i=0}^j \Delta_i)
[g(z,\theta/\rho_{j+1}) - g(z,\theta/\rho_j)],
\end{equation}
where ${\tilde \Delta} = \sum_k {\tilde \Delta}_k$, $\Delta_0=0$,
$\rho_0 = 0$ and $\rho_{\ell +1} = 1$. 

{\it Persistence.\/} Let ${\hat N}={\hat N}(t_w,t_w+t)$ be the number of
distinct {\it backward\/} walkers remaining at time zero from all those
starting at the origin during $(t_w,t_w+t)$.  When ${\hat N}=k$ and
$q<\infty$, the probability of no flips at $0$ in this time interval is
$(1/q)^{k-1}$ and so the persistence probability $F_q(t_w,t_w+t)$ is
$\langle (1/q)^{{\hat N}-1} \rangle$ (for $q = \infty$, $F_\infty = P({\hat
N}=1)$).  In the scaling limit the distribution of ${\hat N}$ converges to
that of $N(1, 1+\theta)$, the number of distinct particles, in the {\it
dual\/} Brownian spanning tree, surviving at time zero from all particles
starting at the origin at all times during $(1, 1+\theta)$.  Writing
$h_k(\theta)$ for $P(N(1, 1+\theta) = k)$, we see that $F_q$ converges to
the aging function
\begin{equation}
\label{eq:qinfty}
f_q(\theta)\,=\,
\cases{\sum_{k=1}^\infty h_k(\theta)(1/q)^{k-1}& if $q<\infty$\cr
h_1(\theta)= P(N(1, 1+\theta)=1)& if $q=\infty$\ .\cr}
\end{equation}

The persistence function $f_\infty(\theta)$ (and hence $h_1(\theta)$) can be
evaluated exactly, thus rederiving a result of \cite{FKR97} by quite
different methods, as follows.  Let $(-X,Y)$ denote the (random) spatial
interval with the same ($q =\infty$) spin value at time $1$ as the
origin. The event $A_{x,y}$ that $X>x$ and $Y>y$ means that the backward
Brownian paths starting at $(-x,1)$ and $(y,1)$ coalesce before time $0$.
Proceeding as in our analysis of (\ref{eq:corr1}), we see that
$P(A_{x,y})=1-\phi\left((x+y)/\sqrt2\,\right)$.  The probability density of
$(X,Y)$ is then $\mu(x,y)=\frac1{2\sqrt\pi}(x+y)e^{-(x+y)^2/4}$ for
$x,y>0$.  Now, given $(X,Y)=(x,y)$, $f_\infty(\theta)$ is the probability
that the {\it forward\/} Brownian paths starting at $(-x,1)$ and $(y,1)$ do
not touch the origin during $(1,1+\theta)$, and thus
\begin{eqnarray}
\label{eq:pers1}
f_\infty(\theta) = 
\int_{0}^\infty\!\!\!\!\int_{0}^\infty\!\!\!\! dx\, dy\,
\mu(x,y)\,\phi\!\left(\frac{x}{\sqrt\theta}\right)
\phi\!\left(\frac{y}{\sqrt\theta}\right).
\end{eqnarray}
After further analysis of the same kind used to derive 
(\ref{eq:corr4})-(\ref{eq:corr5}),
one finds $f_\infty(\theta)=\frac2\pi\arcsin(1/(1+\theta))$
as in \cite{FKR97}.
This formula is consistent with
the $q=\infty$ persistence exponent value of one
\cite{DHP96,DBG94,D95} (for $t_w$ fixed and $t\to\infty$) since
$f_\infty (\theta)$ is asymptotic to $1/\theta$.

{\it Discussion.\/} We presented a powerful and very general
approach, based on coalescing random walks and Brownian paths run forward
{\it and\/} backward in time, to nonequilibrium dynamics in $1D$.  
It yields exact, closed-form expressions in the scaling limit for a
variety of aging (and persistence) quantities including the spin-spin
correlation
$\langle\vec{\sigma}_x(t_w)\cdot\vec{\sigma}_{x'}(t_w+t)\rangle$ for the
$q$-state Potts model for all $q$, following a quench from $T=\infty$ to
$T=0$.  
This type of approach,
based on an exact analysis of the space-time
scaling limit for the entire dynamical process,
should yield
exact expressions for aging functions in a wide
variety of $1D$ systems.

We also presented an exact, closed-form expression for 
the spin-spin
correlation when the system undergoes a sequence of $T=\infty$ interludes.
Perhaps surprisingly, we find that the
effect of such interludes is {\it independent\/} of their timing {\it
provided\/} they occur during the interval $(t_w,t_w+t)$. We believe our
methods may work also for $T<\infty$ interludes, which represent a common
experimental situation \cite{review}, and for other aging quantities
of interest; these analyses will be deferred to a later paper.

{\it Acknowledgments.\/} We thank D.~ben-Avraham, B.~Derrida, and
C.~Doering for helpful correspondence, G.~Biroli for pointing out several
references, and J.L.~Lebowitz for useful remarks.  This research was
partially supported by CNPq PRONEX Grant 662177/1996-7 and Grant
300576/92-7, FAPESP Theme Grant 99/11962-9, Universit\`a di Bari Grant
(es.~fin.~2000), and by NSF Grants DMS-98-02310, DMS-98-03267 and
DMS-98-02153.

\end{document}